\documentclass[prl,twocolumn,superscriptaddress,showpacs]{revtex4}
\usepackage{graphicx}
\usepackage{amssymb}
\begin{document}
\title{Conductance through a One-Dimensional Correlated System: \\
Relation to Persistent Currents and Role of the Contacts}
\author{Rafael A.\ Molina}
\affiliation{Institut de Physique et Chimie des Mat{\'e}riaux de Strasbourg, 
UMR 7504 (CNRS-ULP), 23 rue du Loess, 67037 Strasbourg, France}
\author{Dietmar Weinmann}
\affiliation{Institut de Physique et Chimie des Mat{\'e}riaux de Strasbourg, 
UMR 7504 (CNRS-ULP), 23 rue du Loess, 67037 Strasbourg, France}
\author{Rodolfo A.\ Jalabert}
\affiliation{Institut de Physique et Chimie des Mat{\'e}riaux de Strasbourg, 
UMR 7504 (CNRS-ULP), 23 rue du Loess, 67037 Strasbourg, France}
\author{Gert-Ludwig Ingold} 
\affiliation{Institut f{\"u}r Physik, Universit{\"a}t Augsburg, 
Universit{\"a}tsstra{\ss}e 1, 86135 Augsburg, Germany}
\author{Jean-Louis Pichard} 
\affiliation{CEA/DSM, Service de Physique de l'Etat Condens{\'e}, 
Centre d'Etudes de Saclay, 91191 Gif-sur-Yvette, France}  
\date{July 25, 2002}
           
\begin{abstract}
Based on a recent proposal
[O.P.\ Sushkov, Phys.\ Rev.\ B {\bf 64}, 155319 (2001)], we relate
the quantum conductance through a sample in which electrons are 
strongly correlated 
to the persistent current of a large ring, composed of the 
sample and a non-interacting lead. A scaling law in 
the lead length allows to extrapolate to a well-defined value of the 
conductance, depending only on intrinsic properties of the sample and the 
nature of the contacts between the sample and the lead. For 
strongly disordered samples, the conductance is found to be enhanced 
by the interaction.  
\end{abstract}
\maketitle


Viewing quantum transport as a scattering problem
\cite{Landauer57,Buttiker86} generated a new understanding of 
the electronic conductance. This approach is able to 
explain a wealth of experimental results \cite{Imry} 
in mesoscopic systems when electron-electron (e-e) correlations
are not important. To include these correlations is non-trivial 
and remains one of the major challenges in the field 
(see e.g.\ \cite{Zotos}). While none of the proposals to calculate 
the conductance for a correlated system is well suited for 
numerical calculations \cite{meir,pastawski} or free of certain 
assumptions \cite{berkovits}, such an issue becomes crucial 
in present day's research exploring electronic transport through 
nanosystems (carbon na\-no\-tubes \cite{nanotubes},  
molecules \cite{molecular_electronics}, and point contacts \cite{sushkov}), 
where the Coulomb repulsion leads to important correlations. 

Reservoirs and leads are key elements in the scattering approach, 
and possess very clear physical meanings since the measurement is made 
with electrodes which behave as electron reservoirs. In a good 
electrode, the electron density $n_e$ is large, the ratio $r_s$ between 
Coulomb and Fermi energy is small, hence the e-e interaction is 
negligible.  
In contrast, $n_e$ in a nanosample can be very small, 
yielding a large ratio $r_s$ and important e-e correlations.
 
The dimensionless conductance $g$ does not only depend on the intrinsic 
properties of the sample, but also on the way it is connected to the 
electrodes. The quality of the contacts is particularly important 
for correlated electrons. For a clean Luttinger liquid attached to 
non-interacting leads through reflectionless contacts, it has been found 
\cite{safi} that the interactions do not influence $g$. In the 
other extreme, if the contacts are tunnel barriers, the 
interactions lead to Coulomb blockade \cite{kouwenhoven}, thereby
dominating $g$. In carbon na\-no\-tubes, various transport regimes are 
observed depending on the nature of the contacts \cite{nanotubes}.

As shown by Kohn \cite{kohn} and Thouless \cite{Thouless}, 
$g$ is also related to the sensitivity of the sample's eigenstates to 
a change of the boundary conditions. 
This sensitivity can be tested by closing a system to a ring and measuring 
the persistent current as the response to an enclosed magnetic 
flux $\phi$. At zero temperature, the persistent current is given by 
$J=-\partial E/\partial \phi$, 
where $E$ is the ground state energy of the many-body system. 
Interactions play an important role for $J$, and it is generally 
accepted that they account for the large difference 
between experiments and one-particle calculations \cite{reviewpercurr}.
There have been various attempts \cite{shastry,berkovits} 
to link $J$ and $g$ for an interacting ring. However, the ring 
built from the sample itself does not contain any 
reservoirs in which energy relaxation can take place.  
Negative zero-frequency conductivities occur \cite{fye}, unlike in 
the dissipative case in which we are interested here.

As pointed out in Refs.~\cite{meir,sushkov}, at zero temperature, 
not only for the non-interacting case, but also for correlated samples,
$g$ is given by $|t(E_F)|^2$, the probability for an electron at 
the Fermi energy $E_F$ to be elastically transmitted through 
the sample. 
Moreover, if one replaces the massive 
electrodes (with negligible e-e correlations) used in a real measurement 
by very long non-interacting one-dimensional leads, one can expect 
that they have a similar effect \cite{pastawski}. 
Sushkov recently proposed \cite{sushkov} that $|t(E_F)|^2$ 
can be extracted from the persistent current of a 
much larger ring, composed of the sample itself, together with 
a long lead closing the system. This has the 
considerable advantage that a ground state property ($J$) suffices
to determine $g$. However, one needs the $J$ of the {\it combined
system}\/  (sample plus lead), and not the one of the 
system alone as in previous works \cite{berkovits,Thouless,shastry,kohn}.

In the following, we adapt the approach of Ref.~\cite{sushkov} 
to calculate $g$ for 
one-dimensional interacting electrons using 
the Density Matrix Renormalization Group (DMRG) algorithm 
\cite{white,peter}. We check that a scaling law allows to 
extrapolate to an infinite lead, yielding $g$ as a property of the sample 
and the way it is connected to the lead. The nature of the contacts turns out 
to play a major role. Then, we extend our analysis to disordered samples, 
where we find that, similarly to the case of persistent currents \cite{sjwp}, 
repulsive interactions may increase $g$ for strong disorder.

We first present an alternative derivation 
of Sushkov's result \cite{sushkov}, pointing out the main assumptions, 
for the non-interacting case. As depicted in the upper inset of
Fig.~1, we consider a sample (S, hashed region) closed to a ring 
by a non-interacting and disorder-free lead (L), and threaded by a 
flux $\phi$. The total 
length $L=L_{\rm S}+L_{\rm L}$ consists of the sample length $L_{\rm S}$ 
and the lead length $L_{\rm L}$. The one-particle 
eigenstates of the total system satisfy
\begin{equation}\label{eq:quantcond}
{\rm det}(I-M_{\rm S} M_{\rm L})=0 \, ,
\end{equation}
with the transfer matrices of the sample and the lead
\begin{equation}\label{eq:transfer}
\begin{array}{rcl}
M_{\rm S} &=& {\displaystyle\frac{1}{\sin\varphi}} \left( \begin{array}{cc}
e^{i\alpha}/\sin\theta & -i\cot{\theta}+ \cos{\varphi} \\
i\cot{\theta}+ \cos\varphi & e^{-i\alpha}/\sin\theta
\end{array} \right) \vspace{1mm}\\
M_{\rm L} &=& e^{i \Phi} \left( \begin{array}{cc} 
e^{i k L_{\rm L}} & 0 \\ 
0 & e^{-i k L_{\rm L}}
\end{array} \right) \, ,
\end{array}
\end{equation}
respectively. Here, $\Phi=2 \pi \phi/ \phi_0$ where 
$\phi_0$ is the flux quantum. 
The scattering is characterized by the angle $\theta$,  
the phase-shift $\alpha$, and the angle $\varphi$ (equal to $\pi/2$ 
if right-left symmetry is respected). These angles are functions 
of $k$, the wave-vector in the lead. 
The transmission amplitude is given by 
$t = e^{i\alpha}\sin{\theta}\sin{\varphi}$. With (\ref{eq:transfer}),
the quantizing condition (\ref{eq:quantcond}) can then be written as
\begin{equation}\label{eq:quantcond2}
\cos{\Phi} = \frac{1}{|t|} \, \cos(k L + \delta \alpha) \, ,
\end{equation}
with the relative phase-shift $\delta \alpha= \alpha - k L_{\rm S}$. The 
persistent current carried by a one-particle state (with energy $\epsilon$)
is $j(\phi)=-(\partial \epsilon/\partial k) (\partial \phi/\partial k)^{-1}$. 
We work at $\Phi=\pi/2$ and establish two crucial 
assumptions: i) $|\partial (\delta\alpha)/\partial k| \ll L$; ii) 
$\partial \epsilon/\partial k \simeq \hbar^2k/m$. The first one
states that the Wigner time associated with the scattering region
is negligible compared with the time spent in the leads. Notice
that we work with a relative Wigner time
$\tau_W=(m/\hbar^2k)\partial(\delta\alpha)/\partial k$, that is, 
the difference between the delay time of the scattering region and that 
of a potential-free region having the same length. The second assumption
implies that the dispersion relation is essentially unaffected by the
scattering potential. 

The persistent current of $N$ non-interacting spinless fermions
(for simplicity we take $N$ even) is given by \cite{riedel}
\begin{equation}\label{eq:totpercurr}
J(\Phi=\pi/2) = \frac{e\hbar}{m L} \ k_F |t(k_F)| \, .
\end{equation}
Denoting by $J^0$ the persistent current of a clean ring of length $L$, 
the conductance may be expressed as \cite{sushkov}
\begin{equation}\label{eq:susrel}
g  = \lim_{L_{\rm L}\rightarrow\infty}
\left(\frac{J(\pi/2)}{J^0(\pi/2)}\right)^2 \, .
\end{equation}

 With interaction \cite{stafford}, assumptions i) and ii) always hold in the 
large $L$ limit. Moreover, the use of Eq.~(\ref{eq:susrel}) implies
that the one-particle states of the correlated system can still 
be indexed by the lead wave-vectors $k$. That is, adding an infinite 
non-interacting lead to a finite non Fermi-liquid sample restores the 
Fermi-liquid behavior. This assumption, which has 
been used in the perturbative calculation of transport through
Hubbard chains connected to reservoirs \cite{oguri}, 
requires that the interactions are {\it completely}\/ switched off 
in the one dimensional lead. Otherwise the Luttinger liquid behavior 
\cite{kane_fisher} sets in, and one cannot obtain a result which 
is independent of the length of the auxiliary lead. In this, our 
approach differs from Sushkov's, where the interactions in 
the lead are kept (within the Hartree-Fock approximation).

\begin{figure}
\includegraphics[width=\columnwidth]{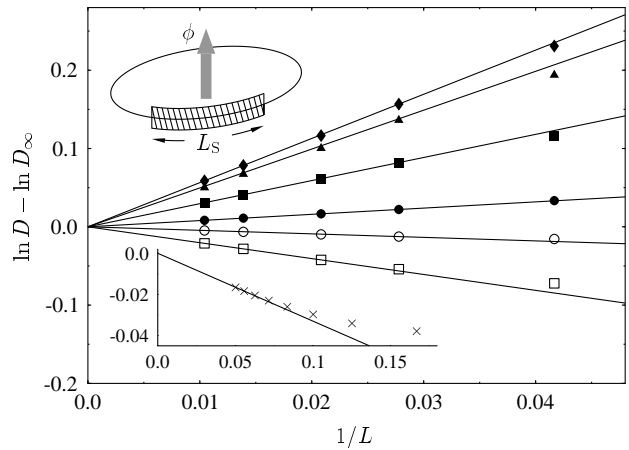}
\caption{Scaling of $\ln D$ with the total length $L$ of the
system, showing a linear increase of $\ln D$ with $1/L$ 
for even $L_{\rm S}=12$ ($U=1$ ($\bullet$), 
$U=2$ ($\blacksquare$), $U=3$ ($\blacktriangle$), $U=4$
($\blacklozenge$)), and a decrease for odd $L_{\rm S}=13$ 
($U=1$ ($\circ$) and $U=2$ ($\square$)). Lower inset: scaling for 
unpolarized electrons within the Hubbard model ($L_{\rm S}=2$, $U=1$). 
Upper inset: sketch of a ring consisting of the sample (hashed region)
and a non-interacting lead threaded by a flux $\phi$.}
\label{fig1}  
\end{figure}

Eq.~(\ref{eq:susrel}) allows to calculate $g$ from 
ground-state energies. We do this for spinless fermions (polarized 
electrons) in a ring described by the Hamiltonian
\begin{equation}\label{eq:hamiltonian}
H = -\sum_{i=1}^{L}(c^\dagger_i c_{i-1} + c^\dagger_{i-1}c_i)
+ \sum_{i=2}^{L_{\rm S}}U\left[n_i-{1\over 2}\right]
\left[n_{i-1}-{1\over 2}\right]
\end{equation}
where $c_i$ ($c^\dagger_i$) is the annihilation (creation)
operator at site $i$, $n_i = c^\dagger_i c_i$ the number
operator, and the flux-dependent boundary condition enters
through $c_0 = \exp(i\Phi )c_L$. The interaction is restricted
to nearest neighbors and effective in the sample, but vanishing in the
lead. It is equilibrated by a compensating potential that prevents
the particles from emptying the interacting region. The form of
the Hamiltonian allows to have particle-hole symmetry at
half filling. We work with a number of fermions $N=L/2$, such that 
the mean density is always $1/2$ independently of $L_{\rm S}$ and 
$L_{\rm L}$. 

Using the DMRG algorithm as described in Ref.~\cite{peter}, 
we calculate the ground-state
energies $E(\Phi)$ at $\Phi=0$ and $\Phi=\pi$, to obtain the
stiffness $D = (L/2)|E(0)-E(\pi)|$ (which is simpler to calculate than
$J$). When the flux-dependence of the ground-state energy is 
sinusoidal, which is the case in one-dimensional localized systems, 
$D$ is directly proportional to the maximum $J$. We will use such 
an identification to calculate $g$ from Eq.~(\ref{eq:susrel}), taking 
$D$ instead of $J$ \cite{stafford}.

An obvious requirement for Eq.~(\ref{eq:susrel}) to be useful is that the 
resulting $g$ should be independent of $L_{\rm L}$, since the lead is 
an auxiliary element of
the procedure. Therefore, the first numerical step is to compute $D$ for 
increasing $L_{\rm L}$ with given $L_{\rm S}$ and $U$ (Fig.~\ref{fig1}). 
We find a very clear asymptotic behavior, described by the scaling law
\begin{equation}\label{eq:scaling}
D(U,L_{\rm S},L_{\rm L}) = D_{\infty}(U,L_{\rm S}) \
\exp{\left(\frac{C(U,L_{\rm S})}{L}\right)} \, ,
\end{equation}
where the {\em intrinsic value} $D_{\infty}$ is {\em independent} 
of the length of the auxiliary lead. This asymptotic value $D_{\infty}$ is 
then used to determine the conductance as $g=(D_{\infty}/D^0_{\infty})^2$,
where $D^0_{\infty}$ corresponds to the clean non-interacting ring.  
The sign of $C$ \cite{tau_W} depends on the parity of the number of sites
$L_{\rm S}$ in the sample: $C>0$ 
for even $L_{\rm S}$ 
(filled symbols) and $C<0$ for odd $L_{\rm S}$ (open symbols).
\begin{figure}
\includegraphics[width=\columnwidth]{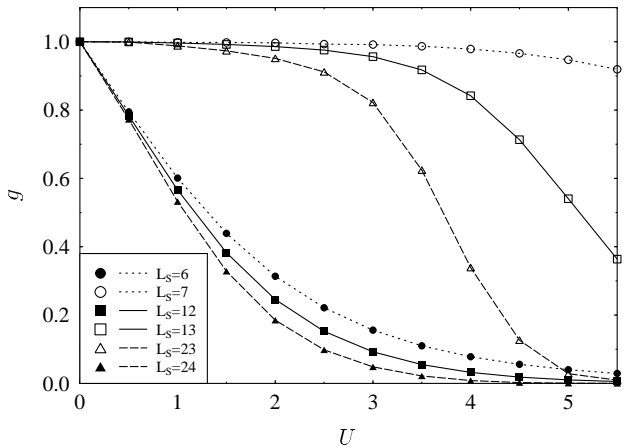}
\caption{Conductance $g$ as a function of the interaction strength $U$
  for different values of the sample length $L_{\rm S}$.}
\label{fig2}
\end{figure}

The method also works for non-polarized electrons 
(Hubbard model with on-site interaction), and the size-scaling again 
allows to obtain intrinsic values (lower inset in Fig.\ \ref{fig1}). 
In the sequel we concentrate on the spinless case 
(Hamiltonian (\ref{eq:hamiltonian})), which contains the main features  
we are interested in, and allows to reach larger samples.
 
Having verified the consistency of our approach, we now
study the systems of interest. In Fig.~\ref{fig2} we present 
the conductance, as a function of $U$, for various 
sample lengths $L_{\rm S}$. One observes a monotonic decrease with increasing 
$U$, and a very clear even-odd asymmetry according to the parity of 
$L_{\rm S}$. Samples with odd $L_{\rm S}$ 
exhibit almost perfect transmission up to large values of
$U$, while an even $L_{\rm S}$ results in a decrease of $g(U)$ already at
weak interaction. For odd $L_{\rm S}$, particle hole symmetry leads to 
degenerate sample configurations with $(L_{\rm S} \pm 1)/2$ particles
in the interacting region. This is similar to a Coulomb blockade
resonance. The traveling particle can thus 
become trapped for a long time ($\tau_W > 0$), consistent with 
negative $C$ \cite{tau_W}. Since the two configurations are coupled by
processes which transfer particles through the interacting sample, 
one obtains perfect transmission, as in the perturbative treatment
of Ref.~\cite{oguri}. 

On the other hand, an even number of sites implies that the transport 
of one particle through the sample takes place via a virtual state with 
an energy of order $U$ above the ground-state. Thus, no resonance
can be expected and the transmission, which is suppressed already by 
moderate $U$, is a fast process with $\tau_W < 0$, consistent with $C
> 0$. In addition, increasing $L_{\rm S}$ reduces $g$ linearly for small
$U$, and exponentially for $U>2$, consistent with the
Mott-insulating behavior.

\begin{figure}
\includegraphics[width=\columnwidth]{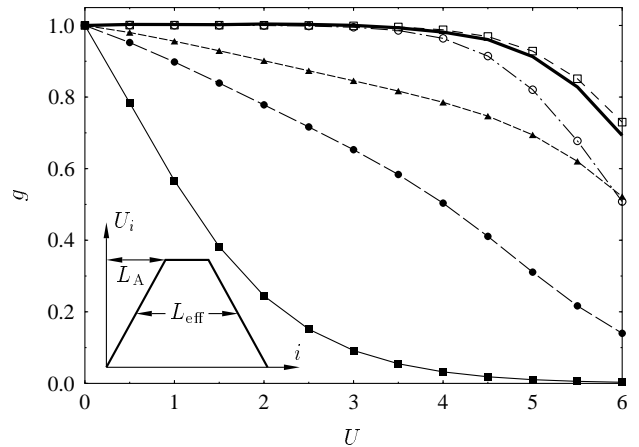}
\caption{Conductance $g$ as a function of the interaction strength
$U$, for a fixed $L_{\rm eff}=12$, and increasing smoothing of the
contacts, defined by the length $L_{\rm A}$ (see inset). 
Filled (open) symbols correspond to even (odd) $L_{\rm S}$;
$L_{\rm A}=1$ ($\circ$), and $L_{\rm A}=9$ ($\square$), 
$L_{\rm A}=0$ ($\blacksquare$), 
$L_{\rm A}=4$ ($\bullet$), and $L_{\rm A}=10$ ($\blacktriangle$). 
Using the same smoothing length ($L_{\rm A}=10$) but improving in the
shape (a $\tanh$ function (thick solid line) instead of a linear
increase ($\blacktriangle$)) helps $g$ to approach the
perfect value.}
\label{fig3}
\end{figure}
The even-odd asymmetry, and the perfect transmission for the odd case,
point to the importance of the contacts. In order to investigate their 
role, we introduce a position dependent interaction strength $U_i$ 
which increases linearly from $0$ to its maximum value $U$, inside 
the ``contacts'' of length $L_{\rm A}$ (see inset of Fig.~\ref{fig3}). 
As shown in Fig.~\ref{fig3}, these smooth contacts increase $g$, at 
constant effective length $L_{\rm eff}=L_{\rm S}-L_{\rm A}$ of the
sample. The effect is pronounced in the case of even $L_{\rm S}$ and the 
conductance approaches the ideal situation of perfect transmission 
expected for reflectionless contacts when we improve the smoothing 
\cite{safi}. 
While the form of the contacts preserves the
right-left symmetry, the smoothing leads to an extension of the perfect 
transmission at odd $L_{\rm S}$ towards higher values of $U$.
However, we found that asymmetric contacts destroy the perfect 
transmission. The
strong influence of the contacts is crucial when describing
experiments since it seems impossible to connect a nanosample 
via reflectionless contacts. It also shows the limitation of other 
approaches relating the conductance of an interacting sample to its 
intrinsic properties, without taking into account the way it is 
connected to the electrodes.  

\begin{figure}
\includegraphics[width=\columnwidth]{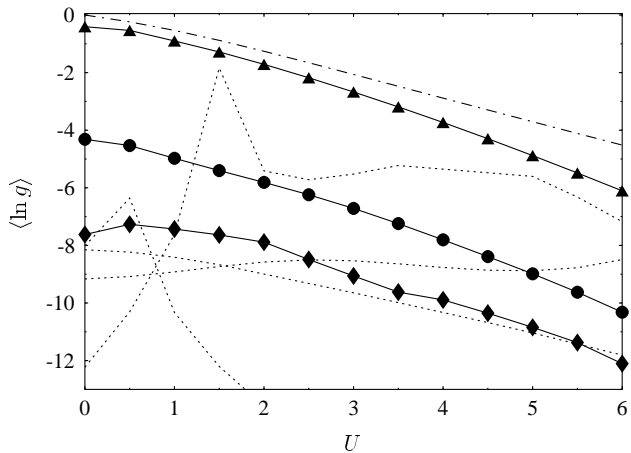}
\caption{Ensemble averages of $\ln g$ as a function of the
  interaction at $L_{\rm S} = 8 $, in the presence of disorder $W=1$
($\blacktriangle$), $5$ ($\bullet$) and $9$ ($\blacklozenge$).
The statistical errors are smaller than the symbol size. 
The dash-dotted line represents the clean case, the dotted
 lines represent four individual samples at $W=9$.}
\label{fig4}
\end{figure}
While clean interacting systems are of physical interest ({\it e.g.}\ 
carbon na\-no\-tubes), it is also important to consider the generic case of
disordered systems. To this end, we add the term 
$W\sum_{i=1}^{L_{\rm S}}v_i n_i$ to the Hamiltonian
(\ref{eq:hamiltonian}), where $W$ is the disorder strength 
and the $v_i$ are distributed equally in $[-1/2,1/2]$. 
We have checked that $D$ scales with $L$ as before 
(Eq.~(\ref{eq:scaling})), ensuring a
well-defined limiting value for $g$. The 
even-odd dependence of $g$ disappears when disorder is introduced. 
The combined effect of disorder and interactions on $g$ 
is shown in Fig.~\ref{fig4}. In the ballistic regime (at $W=1$ 
the mean free path exceeds $L_{\rm S}$) the effect of $W$ is weak at
small $U$, and it becomes more pronounced at stronger $U$ (when
the disorder pins the Mott insulator, reinforcing the 
localization). At large $W$, $g$ for individual samples exhibits 
peaks as a function of $U$ whenever a charge reorganization 
occurs, similarly to the case of the $D$ calculated without the 
lead \cite{sjwp}. 
Very remarkably, the ensemble average of $\ln g$ is {\em increased} by a
moderate repulsive $U$, showing the non-trivial interplay of 
disorder and interactions in a transport problem. 


In conclusion, starting from a recent proposal \cite{sushkov}, 
we have provided a well-defined procedure 
for calculating the conductance $g$ of interacting
one-dimensional wires, and used it to investigate 
correlation and disorder effects. While the interaction 
reduces $g$ for spinless fermions in the presence of weak 
or moderate disorder, a moderate repulsive interaction 
increases $g$ at strong disorder.      
We also determined the crucial role of the sample-to-lead
contacts on the conductance. 


We thank O.\ Sushkov and C.\ Stafford for useful discussions, 
P.\ Schmitteckert and Ph.\ Brune for their DMRG programs. 
RAM acknowledges the financial support from the European 
Union's Human Potential Program (contract HPRN-CT-2000-00144).  
RAJ and DW thank the INT at the University of Washington for 
its hospitality and support during completion of this work. 



\begin{thebibliography}{99}
\bibitem{Landauer57}
R.\ Landauer, IBM J.\ Res.\ Dev.\ {\bf 1}, 223 (1957).  
\bibitem{Buttiker86}
M.\ B\"uttiker, Phys.\ Rev.\ Lett.\ {\bf 57}, 1761 (1986).
\bibitem{Imry}
Y.\ Imry, {\it Introduction to Mesoscopic Physics}, 
Oxford University Press (New York 1997).
\bibitem{Zotos}
P.\ Prelovsek and X. Zotos, cond-mat/0203303.
\bibitem{meir}
Y.\ Meir and N.S.\ Wingreen, Phys.\ Rev.\ Lett.\ {\bf 68}, 2512 (1992).
\bibitem{pastawski}
H.M. Pastawski, Phys.\ Rev.\ B\ {\bf 44}, 6329 (1991).
\bibitem{berkovits}
R.\ Berkovits and Y.\ Avishai, Phys.\ Rev.\ Lett.\ {\bf 76}, 291 (1996).
\bibitem{nanotubes} 
P.L.\ McEuen, M.S.\ Fuhrer, and H.\ Park,  
IEEE Trans.\ Nanotech.\ {\bf 1}, 78 (2002), and references therein.
\bibitem{molecular_electronics} see {\it e.g.}\ Special issue 
on {\it Transport in Molecular Wires}, ed.\ by 
P.\ H{\"a}nggi, M.\ Ratner, and S.\ Yaliraki,
Chem.\ Phys.\ {\bf 281}, pp.\ 111--487 (2002).
\bibitem{sushkov}
O.P.\ Sushkov, Phys.\ Rev.\ B {\bf 64}, 155319 (2001).
\bibitem{safi}
I.\ Safi and H.J.\ Schulz, Phys.\ Rev.\ B\ {\bf 52}, R17040 (1995).
\bibitem{kouwenhoven} L.\ Kouwenhoven {\it et al.}, 
in {\it Mesoscopic Electron Transport}, 
ed.\ by L.L.\ Sohn {\it et al.} (Kluwer 1997).
\bibitem{kohn}
W.\ Kohn, Phys.\ Rev.\ {\bf 133}, A171 (1964). 
\bibitem{Thouless}
D.J.\ Thouless, Phys.\ Rev.\ Lett.\ {\bf 39}, 1167 (1977).
\bibitem{reviewpercurr}
U.\ Eckern and P.\ Schwab, Adv.\ Phys.\ {\bf 44}, 387 (1995). 
\bibitem{shastry}
B.S.\ Shastry and B.\ Sutherland, Phys.\ Rev.\ Lett {\bf 65}, 243 (1990).
\bibitem{fye}
R.M.\ Fye {\it et al.}, 
Phys.\ Rev.\ B\ {\bf 44}, 6909 (1991).
\bibitem{white} S.R.\ White,
               Phys.\ Rev.\ B {\bf 48}, 10345 (1993).
\bibitem{peter} P.\ Schmitteckert, Ph.D.\ thesis, 
               Univ.\ Augsburg, 1996.
\bibitem{sjwp} 
P.\ Schmitteckert {\it et al.},
Phys.\ Rev.\ Lett.\ {\bf 81}, 2308 (1998).
\bibitem{riedel}
H.-F. Cheung {\it et al.}, 
Phys.\ Rev.\ B {\bf 37}, 6050 (1988).
\bibitem{stafford} 
For a quantum dot in the Coulomb blockade 
regime
[C.A.\ Stafford, R.\ Kotlyar, and S.\ Das Sarma, Phys.\ Rev.\ B {\bf 58}, 
7091 (1998)], 
one obtains $g\propto J^2(\pi/2)$ and $g \propto D^2$.  
\bibitem{oguri}
A. Oguri, Phys.\ Rev.\ B {\bf 59}, 12240 (1999).
\bibitem{kane_fisher}
C.L.\ Kane and M.P.A.\ Fisher, Phys.\ Rev.\ Lett {\bf 68}, 1220 (1992).
\bibitem{tau_W}
In the non-interacting derivation of (\ref{eq:susrel}), 
the first correction in $1/L$ and $\tau_W$ have
opposite sign. A potential well accelerates the particles 
leading to $\tau_W < 0$, and $C>0$. 

\end{thebibliography}
\end{document}